\documentclass[12pt]{article} 

\usepackage{algorithm}
\usepackage{algorithmic}
\usepackage{subcaption}
\usepackage{tabularx}
\usepackage{adjustbox}
\usepackage{float}
\usepackage{placeins}
\usepackage{booktabs}

\begin{document}
	\title{Prediction of ESG Compliance using a Heterogeneous Information Network}	
	\author{
		Ryohei Hisano \\
		Social ICT Research Center \\
		Graduate School of Information Science and Technology \\
		The University of Tokyo\\
		\and
		Didier Sornette \\
		ETH Z\"{u}rich \\
		Department of Management Technology and Economics
		\and
		Takayuki Mizuno \\
		National Institute of Informatics
	}
	\maketitle
	\thispagestyle{empty}
	
\begin{abstract}
	
Negative screening is one method to avoid interactions with inappropriate entities. For example, financial institutions keep investment exclusion lists of inappropriate firms that have environmental, social, and government (ESG) problems. They create their investment exclusion lists by gathering information from various news sources to keep their portfolios profitable as well as green. International organizations also maintain smart sanctions lists that are used to prohibit trade with entities that are involved in illegal activities. In the present paper, we focus on the prediction of investment exclusion lists in the finance domain. We construct a vast heterogeneous information network that covers the necessary information surrounding each firm, which is assembled using seven professionally curated datasets and two open datasets, which results in approximately 50 million nodes and 400 million edges in total. Exploiting these vast datasets and motivated by how professional investigators and journalists undertake their daily investigations, we propose a model that can learn to predict firms that are more likely to be added to an investment exclusion list in the near future.  Our approach is tested using the negative news investment exclusion list data of more than 35,000 firms worldwide from January 2012 to May 2018.  Comparing with the state-of-the-art methods with and without using the network, we show that the predictive accuracy is substantially improved when using the vast information stored in the heterogeneous information network.  This work suggests new ways to consolidate the diffuse information contained in big data to monitor dominant firms on a global scale for better risk management and more socially responsible investment.

\end{abstract}

\section{Introduction}

Negative screening is one method to avoid interactions with inappropriate entities. For instance, international organizations and governments issue smart sanctions lists to prohibit trade with foreign entities that are involved in illegal activities, such as terrorism and money laundering. Financial institutions also maintain many versions of investment exclusion lists by gathering information from various news sources. Their focus is not only to avoid firms that have financial problems to keep their portfolios profitable, there is also a growing interest to put pressure on publicly listed firms to improve their environmental, social, and government (ESG) practices \cite{OECD2017}, that is, to put more pressure on big firms ``to do the right thing'' by avoiding investing in them. The aims of these ESG practices include not only environmental issues but also human rights issues (e.g., child labor), discrimination (e.g., gender and race) issues, and incorporating information from smart sanctions lists issued by countries and international organizations worldwide. Thus, negative screening is becoming increasingly important to enhance the healthy functioning of global markets.

\begin{figure*}[ht]
	\centering
	\begin{subfigure}{.3\textwidth}
		\centering
		\includegraphics[width=1.0\linewidth]{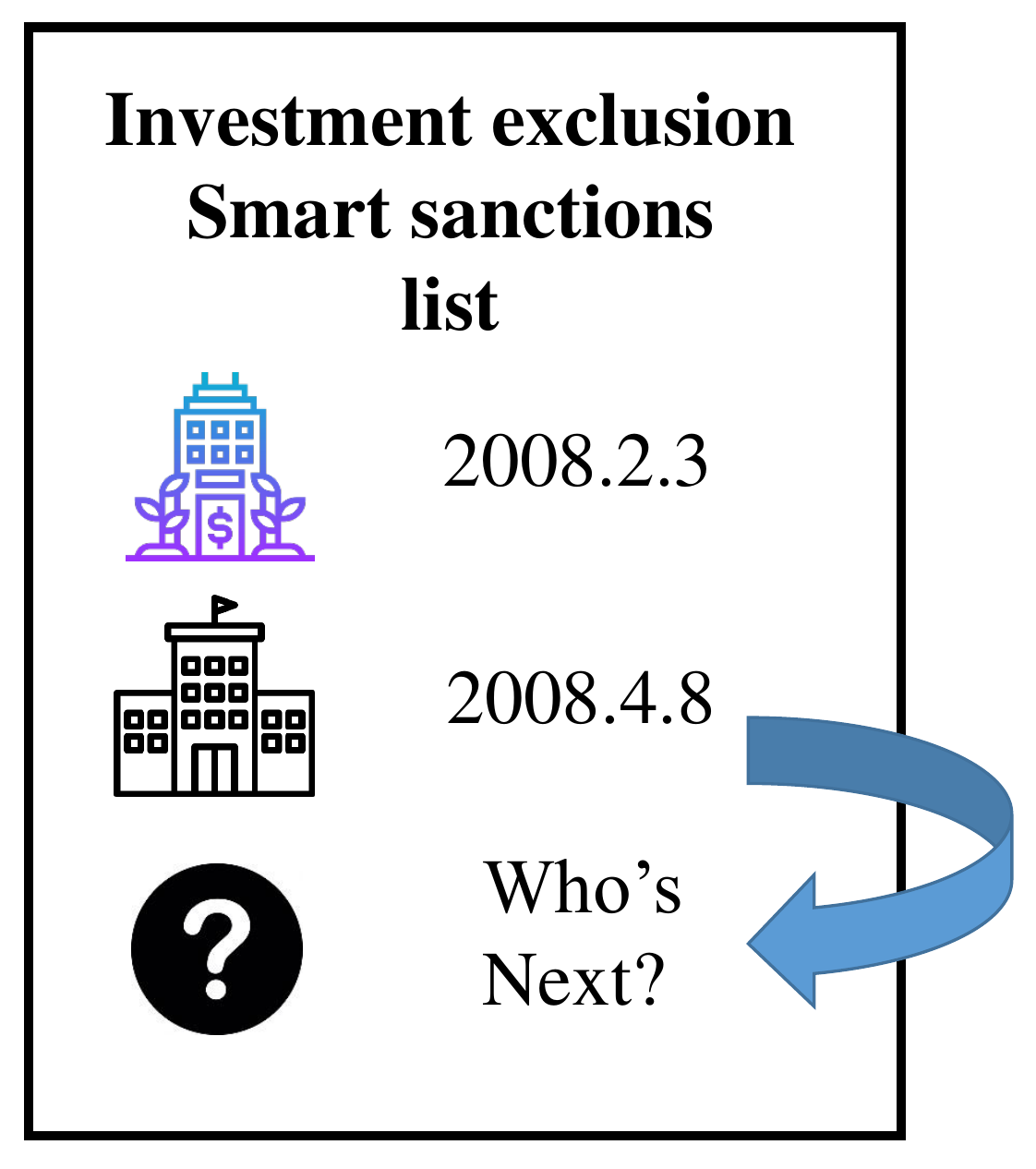}
		\caption{List prediction problem.}
		\label{fig:list}
	\end{subfigure}%
	\begin{subfigure}{.7\textwidth}
		\centering
		\includegraphics[width=1.0\linewidth]{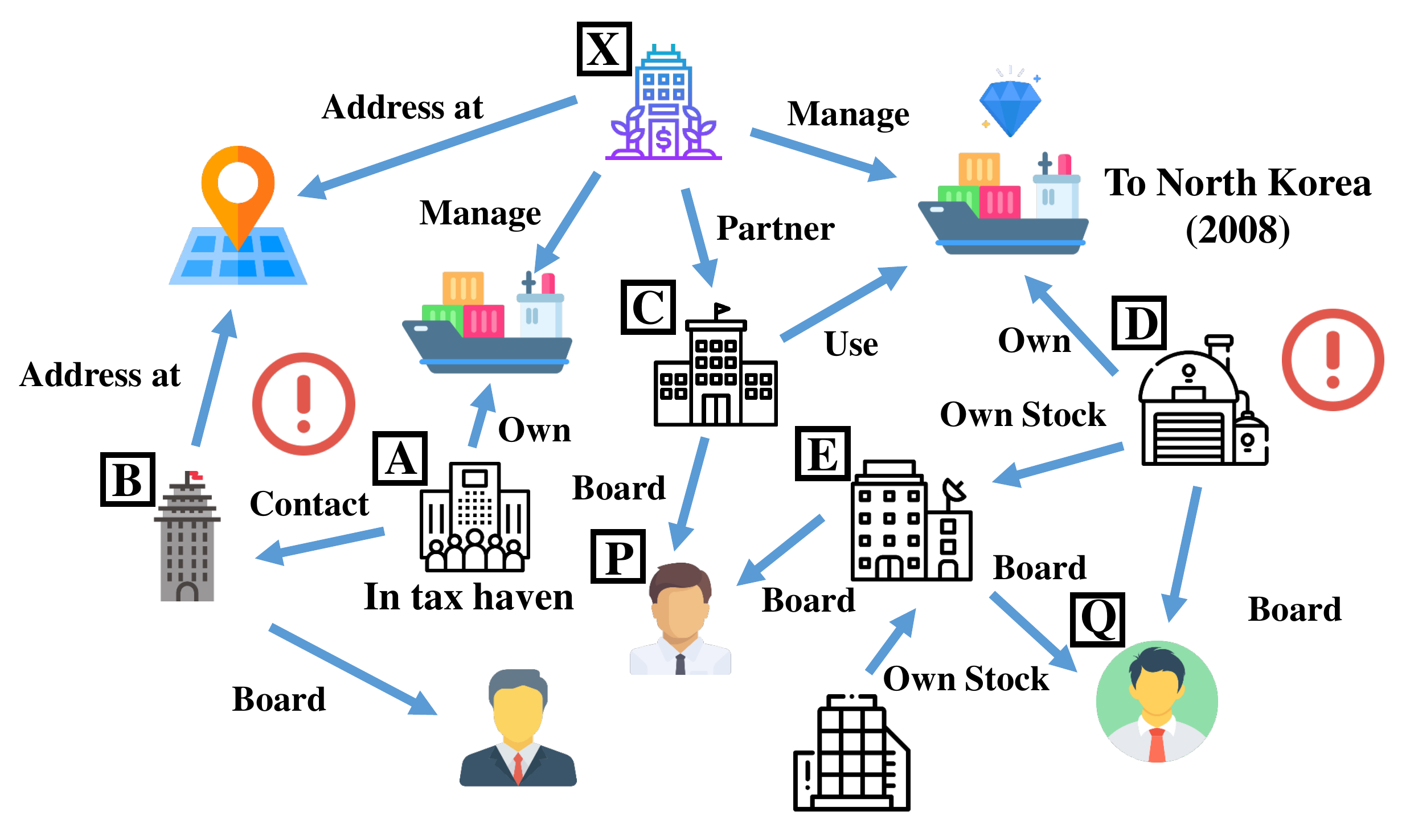}
		\caption{Simplified network that illustrates the investigation.}
		\label{fig:investigation}
	\end{subfigure}%
	\caption{Schematic figures describing the problem.}
	\label{fig:schematic}
\end{figure*}

Our focus is precisely to predict the appearance of firms on investment exclusion lists maintained in the finance domain (Fig.~\ref{fig:list}), which is gaining popularity worldwide \cite{Sherwood2018}.  There are three information sources used to create such investment exclusion lists: (1) information that firms voluntarily disclose, (2) ESG ratings provided by rating agencies and (3) news information reported by the media. We focus on the investment exclusion lists created using news information because (1) is susceptible to manipulation, as in the Enron's creative accounting practices \cite{Markham2006}, and (2) might be corrupted by conflicts of interest, as in the subprime mortgage crisis \cite{Hill2010}. Although there are concerns about fake news, news reportings used for professional investments are less susceptible to manipulation, and investment exclusion lists created from these news reportings are widely reported to have a positive impact on a portfolio's performance \cite{Sherwood2018}.  However, news information also has its shortcomings, such as investors could react only after the news is released. Their ex-post nature makes them effectively ``locking the barn door after the horse has been stolen.". A more ambitious approach is to try to identify possible future news events that have not yet been reported that might trigger a firm to be added to the investment exclusion lists, as we propose here.

Our approach is tested using negative news investment exclusion list data of more than 35,000 firms worldwide from January 2012 to May 2018.  Our investment exclusion lists are based on data from Dow Jones, which created its dataset using negative news information from about 10,000 news sources worldwide.  Dow Jones categorizes negative news into 17 categories, and we create investment exclusion lists according to this classification.  Because the strategy to predict firms that might be exposed to a financial problem in the near future might be different from the strategy to predict firms that might be exposed to environmental problems, we must have a method that can adjust its prediction strategy to each investment exclusion list category accordingly.  Thus, we aim to build a model that can adaptively adjust to each category.

However, it is not sufficient to develop an adaptable prediction strategy for each investment exclusion list category by using only basic information that one data vendor provides (i.e., date of addition, industry classification, and headquarters location).  Thus, we construct a vast heterogeneous information network that covers the necessary information surrounding each firm by gathering information from several sources.  The network is assembled using seven professionally curated datasets and two open datasets, which results in approximately 50 million nodes and 400 million edges in total.  Exploiting this vast heterogeneous information network, we propose a model that can navigate through the network to predict firms that are more likely to be added to each investment exclusion list in the near future.

To further motivate the heterogeneous information network approach in our setting, we provide a specific example of how real investigators and journalists solve the problem of determining possible entities to add to the smart sanctions lists or investigation targets.  This example is from a book written by a former member of the United Nations Panel of Experts on Sanctions Against North Korea \cite{Furukawa2017}.  The Panel of Experts is in charge of the investigation to determine possible candidates to include in the United Nation's smart sanctions lists.  In Fig.~\ref{fig:investigation}, we provide a simplified network that illustrates how the expert conducted his investigation.

In 2008, the Japanese police force exposed one firm, called X, that was attempting to export luxury goods from Japan to North Korea  (Fig.~\ref{fig:investigation}).  The export of luxury goods to North Korea is against United Nations sanctions and thus is illegal in Japan.  It is worth emphasizing that only adding firm X to the smart sanctions list was not sufficient to ban all the illegal export activities.  There could have been other firms involved in these illegal exports, and the goal was to include all of them.  This motivated the expert to investigate further.  Firm X was said to manage several other vessels, one of which was held by a firm in a tax haven (i.e., firm A).  This company's contact information was directed to firm B, which interestingly had the same registered address as company X.  This raised suspicion of these firms (i.e., firm A and B) and further supporting investigations were performed.

In 2008, the Japanese police force exposed one firm, called X, that was attempting to export luxury goods from Japan to North Korea  (Fig.~\ref{fig:investigation}).  The export of luxury goods to North Korea is against United Nations sanctions and thus is illegal in Japan.  It is worth emphasizing that only adding firm X to the smart sanctions list was not sufficient to ban all the illegal export activities.  There could have been other firms involved in these illegal exports, and the goal was to include all of them.  This motivated the expert to investigate further.  Firm X was said to manage several other vessels, one of which was held by a firm in a tax haven (i.e., firm A).  This company's contact information was directed to firm B, which interestingly had the same registered address as company X.  This raised suspicion of these firms (i.e., firm A and B) and further supporting investigations were performed.

Furthermore, firm X had a partnership with firm C, which was using the vessels that were involved in the 2008 arrest.  These vessels were owned by firm D, which raised suspicion that firm D was possibly also heavily involved in the illegal activities.  Initially, the expert also thought of the possibility that firm D was involved just by accident.  However, it turned out that partnership firm C had person P as its board member, who owned another firm, E, of which one of the principal shareholders was firm D, which was under suspicion.  Moreover, firm D and firm E happened to have the same board member, Q, which further reinforced this suspicion.

As is clear from this example, investigators and journalists attempt to track suspicious patterns by manually inspecting information from several sources (i.e., in this case, vessel information, shareholder information, firm relational information, and registry information) to narrow down their list of targets.  However, investigating each entity, manually as the expert above, might not be a reasonable approach when we have a large number of entities to monitor.  Specifically, in the finance domain, there are cases when we need to invest on a global scale for a more diversified portfolio.  There were 46,583 officially listed domestic firms worldwide in 2017 \cite{WFE2017}, and monitoring them on a global scale undoubtedly requires the development of machine-assisted methods.  This requirement motivates us to develop our machine-assisted heterogeneous information network approach.

Many studies exist in data mining regarding building a heterogeneous information network by gathering information from various sources \cite{Hofmann2017}.  Recent prominent work includes that of Google \cite{Dong2014} and Wikipedia's DBpedia \cite{Auer2007}, which are used for search engine optimization.  Using web-based data, these databases are expanding rapidly.  Some researchers even claim that the knowledge graph should be the default data model for learning heterogeneous knowledge \cite{Wilcke2017}.  In recent years, there has been a wide variety of both theoretical ~\cite{Sun2013,Wang2018} as well as applied research ~\cite{Chen2016,Cao2018} that focuses on using a heterogeneous information network.  See \cite{Nickel2016, Wang2017Review} for excellent overviews.  There are also studies that focus on using information from multiple (multimodal) sources not limiting to heterogeneous information network structure \cite{Hu2018}.  However, the entire social impact of such an approach is yet to be known.  Our work is another line of applied research that follows this trend to show that information concerning firms worldwide can be mapped into one heterogeneous information network, and a machine-assisted method can learn patterns that can predict the occurrence of firms appearing in investment exclusion lists maintained by professional institutions.

Our contribution is summarized as follows.

\begin{itemize}
	\item We propose a new social impact problem called list prediction using heterogeneous information network that has a significant impact on risk management and ESG investing \cite{Sherwood2018}.
	\item We propose a new model based on label propagation that could exploit the heterogeneous information stored in the network to answer the list prediction problem.
	\item We tested our models using a real-world vast heterogeneous information network that was assembled using seven professionally curated datasets and two open datasets, resulting in a total of approximately 50 million nodes and 400 million edges.  Our investment exclusion lists are based on negative news stories from January 2012 to May 2018 and cover 35,000 firms around the globe.  We thus believe that this dataset is sufficient to judge the validity of our approach in real-world settings.
	\item Comparing with the state-of-the-art methods with and without the network, we show that the predictive accuracy is substantially improved when using our model with heterogeneous information.
	\item Not only does our model performs well in terms of predictive accuracy, but our model is also interpretable.
\end{itemize}


The remainder of the paper is organized as follows.  In the next section, we briefly provide an overview of our datasets, which we use throughout the paper.  We first review our negative news investment exclusion list data.  We also present direct observations that show that negative media coverage has an impact on financial returns, thereby highlighting the importance of performing such predictions.  We then describe all the datasets used in the paper to create our heterogeneous information network.  In the model section, we describe the model used in this paper.  We first describe our proposed model, which is a variation of label propagation using Jacobi iteration with edge weight learning.  We then describe how to define the features for each edge using information in our heterogeneous information network.  We also describe other state-of-the-art methods with and without using heterogeneous information.  In the result section, we summarize the results.  We show that our method substantially outperforms other methods.  We then discuss the interpretability of our model.  In the final section, we conclude the paper.

\section{Datasets}
\subsection{Negative News Investment Exclusion List}

\begin{table}
	\centering
	\begin{tabular}{lrr}
		Date & Name	& Negative News Category  \\
		\hline
		Jan 3, 2012 & Firm A & Management \\
		Jan 3, 2012 & Firm B & Product/Service \\
		Jan 10, 2012 & Firm C & Regulatory \\
		Jan 11, 2012 & Firm D & Workplace \\
		\hline
	\end{tabular}
	\caption{Sample of the Dow Jones Adverse Media dataset.}
	\label{table:adme}
\end{table}

We use Dow Jones Adverse Media Entity data from January 2012 to May 2018 as our primary data.  The data consist of the name of the firm, date of the news report, and 17 categories that classify the negative news report. Table~\ref{table:adme} shows a sample of the dataset.

In Table~\ref{si:table:count}, we present the number of firms in each category for the 35,657 firms analyzed in this study from January 2012 to May 2018.  ``No. of news stories'' denotes the total number of negative news stories for a particular investment exclusion list category.  ``Unique firms'' denotes the total number of unique firms tagged with a particular piece of negative news at least once.  In the table, ``No. of news stories'' is sometimes much higher than ``Unique firms,'' which indicates that some firms are tagged with the same negative news report category multiple times. When we create our investment exclusion lists, we add each firm to the lists for the date of the initial news report.  We also keep a record of the last date of the news report to determine whether there is an ongoing news report.  We can see that, in addition to financial and environmental issues, there are other investment exclusion list categories, such as ``Product/Service,'' which records negative news, such as drug test failure and recall incidents, and ``Regulatory,'' which records when a firm is reported to have problems with regulatory issues.

\begin{table*}
	\centering
	\resizebox{1.0\textwidth}{!}{
	\begin{tabular}{lrrrrrrr}
		Group & No. of Samples	& 0.01 & 0.05 & 0.5 & 0.95	& 0.99 & Skewness \\
		\hline
		With news & 8685 & -0.233 & -0.102 & 0.005 & 0.098 & 0.191 & -6.521 \\
		Without news & 1667616 & -0.218 & -0.109 & 0.005 & 0.110 & 0.207 & 0.165 \\
		\hline
	\end{tabular}}
	\caption{Comparison of 10 trading day log returns with and without news events.  Numbers in the first row indicate quantiles.}
	\label{table:return}
\end{table*}

\begin{table}[!htp]
		\centering
		\resizebox{0.8\textwidth}{!}{
		\centering
		\begin{tabular}{lrr}
			\hline
			Category &  No. of news stories  &  Unique firms\\
			\hline
			Product/Service &      20,637 &        8,779 \\
			Regulatory &      21,652 &        7,552 \\
			Financial &      22,754 &        3,310 \\
			Fraud &      14,489 &        3,997 \\
			Workforce &       7,523 &        3,963 \\
			Management &      11,220 &        4,063 \\
			Anti-Competitive &       7,748 &        3,620 \\
			Information &       6,401 &        2,873 \\
			Workplace &       6,827 &        2,492 \\
			Discrimination-Workforce &       6,477 &        2,426 \\
			Environmental &       4,083 &        1,887 \\
			Ownership &       4,124 &        2,615 \\
			Production-Supply &       2,878 &        1,869 \\
			Corruption &       3,621 &        1,578 \\
			Human &        496 &         302 \\
			Sanctions &        254 &         157 \\
			Association &        247 &          90 \\
			\hline
		\end{tabular}
	}
	\caption{Number of negative news reported from January 2012 to May 2018 among the 35,657 firms investigated in this study.  ``No. of news stories'' represents the total number of negative news stories for a particular negative news category.  ``Unique firms'' represents the total number of unique firms tagged with a particular negative news category.}
	\label{si:table:count}
\end{table}

To highlight the importance of predicting which firms appear in such a dataset, we first tested whether a negative news report had a financial impact by checking its relationship with a cross-section of returns  using the following steps.  For all US stocks in the dataset, we gathered their prices from January 2012 to May 2018:  there were 1,139 such stocks in total.  For each date in the negative news dataset, we used a 10-day window centered on a specified date.  We then calculated the log return between the start and end dates of the 10-day window, and compared these returns with the 10 trading day log returns outside the window.

Table~\ref{table:return} compares the distributions of stock returns with and without negative news reports.  The quantiles and skewness show that the negative tail of the log-return distribution is more stretched than the positive tail, which agrees with previous studies that argued that negative information has a negative impact on financial returns.  We also performed a two-sample Kolmogorov--Smirnov test for the null hypothesis that the two distributions are from the same distribution.  This was rejected with a $p$-value below $10^{-6}$.

\begin{table*}
	\centering
	\resizebox{\textwidth}{!}{
		\centering
		\begin{tabular}{lrrrrr}
			Source	& Date of acquisition &	Node types & Relation types & No. of nodes	& No. of edges \\
			\hline
			Dow Jones Adverse Media Entity & Dec 2016 & Firm & Location, Homepage & 132,127 & 390,320 \\
			Dow Jones State-Owned Companies & Dec 2016 & State-owned firms	& VIP, Employee, Owner &	280,995	& 702,172 \\
			Dow Jones Watchlist	& Dec 2016 & VIPs, specially interested person & Social relations & 1,826,273 & 8,322,560 \\
			Capital IQ Company Screening Report	& Dec 2016 & Firms & Buyer-seller, borrower etc & 505,789 & 2,916,956 \\
			FactSet	& Dec 2015 & Firm, goods, industry	& Parent-child firm, Issue Stock & 613,422 & 8,213,225 \\
			FactShip & Jan 2017 &	Firm, goods, invoice etc &	Overseas trade etc & 16,137,550 & 36,345,381 \\
			Reuters Ownership & Dec 2016 &	Owners, stocks & Issue, own	 & 1,560,544 & 121,769,151 \\
			Panama papers & Jan 2017 & Entities, officers & Shareholder of, director of & 888,630 & 1,371,984 \\
			DBpedia	& Apr 2016	& Various & Various & 35,006,127 & 249,429,771 \\
			\hline
	\end{tabular}}
	\caption{Summary of the dataset used in this study.}
	\label{table:datasets}
\end{table*}


\subsection{Heterogeneous Information Network}

In addition to negative news information, the Dow Jones Adverse Media Entity data contains basic information about the location and domain information of each firm.  However, this information is not sufficient to predict investment exclusion lists.  Hence, our strategy is to assemble data from other widely used professionally curated sources in the form of a heterogeneous information network. Table~\ref{table:datasets} summarizes the dataset used in the paper.

We note several points about the data.  First, to remove duplicates when combining node information from several sources, we did not only consider the name of the firm.  In addition to name similarity, we determined two firms from different datasets to be the same if any of the following information was precisely the same: (i) their homepage information, (ii) the longitude and latitude information of their addresses, or (iii) their stock symbol.  We manually inspected our strategy and found that it led to a small number of ``false positive'' errors (i.e., incorrectly identifying different nodes as duplicates), but to a large number of ``false negative'' errors (i.e., missing nodes that are duplicates).  This was because we could not remove duplicate firms that did not have a homepage, address, or stock symbol information.  For the sake of robustness check of our results, we tested with several variations of this strategy varying the parameters that govern name similarity and excluding either (i), (ii) or (iii) and found that all of them provides similar results as the one described in the present paper.

\begin{table}[thbp]
	\centering
	\begin{tabular}{lrr}
		\hline
		Rank &                                           Relation &  Number \\
		\hline
		1 &                                         located\_in &  2,723,162 \\
		2 &                                           customer &   717,019 \\
		3 &                                           supplier &   713,434 \\
		4 &                                       own\_stock &   493,316 \\
		5 &                                 belongs\_to\_industry &   359,425 \\
		6 &                       strategic\_alliance &   348,352 \\
		7 &                                 creditor &   339,184 \\
		8 &                                      receive\_goods &   330,311 \\
		9 &                                         send\_goods &   319,292 \\
		10 &                                              issue\_stock &   187,498 \\
		11 &                                      make\_products &   181,574 \\
		12 &                                 competitor &   174,487 \\
		13 &                                    part\_of\_industry &   172,621 \\
		14 &                                 borrower &   153,203 \\
		15 &                                           domain &   131,153 \\
		16 &                                        distributor &   116,262 \\
		17 &                               subsidiary &   107,119 \\
		18 &                           parent-company &   107,117 \\
		19 &       associated-person &   100,699 \\
		20 &                             international\_shipping &    95,050 \\
		21 &                                associate &    72,685 \\
		22 &                                 landlord &    62,904 \\
		23 &                  http://dbpedia.org/ontology/party &    55,653 \\
		24 &                                 employer &    47,901 \\
		25 &                                 employee &    47,184 \\
		\hline
	\end{tabular}
	\caption{
		Tthe top 25 relation types. 
	}
	\label{table:relation}
\end{table}

Second, half of the relational information in our datasets does not include a timestamp.  This is problematic in the sense that it is difficult to ensure that no future information is used when we perform our prediction.  To avoid any information from the future contaminating our heterogeneous information network and to achieve an exemplary evaluation, we only predict future occurrences of negative news after February 1, 2017,  which is after the latest date for which we acquired data (Table~\ref{table:datasets}).  Finally, for the relational information in the Dow Jones Adverse Media Entity dataset, we use the December 2016 version and update only the negative news information to May 2018.

We also removed relation types that appeared too many times in our dataset to avoid computational overload.  These relation types include ``http:// dbpedia.org/ontology/wikiPageWikiLink'' and ``http://purl.org/dc/terms/subjects,'' which create approximately 175 million and 22 million edges, respectively.  We also ignored relation types that only appeared in the dataset fewer than 100 times.  Furthermore, some of the edges in our dataset had multiple timestamps, and we unified them into one relationship.  These include relation types such as ``own stocks'' and ``sends goods,'' of which the former are on a quarterly basis, whereas the latter includes the timestamp information of when they passed through US customs.  For ``own stocks,'' we further restricted the data to relationships with at least 5\% ownership.  After the removal of duplicates and data cleaning, a total of approximately 3.7 million nodes and 9.1 million edges with 216 relation types remained.  Table~\ref{table:relation} shows the top 25 relation types in our dataset.  Many relation types connect the firms, but there are also relation types, such as those for (i) associations and employees, which relate firms to people; (ii) own stocks, which relates firms or individuals to a stock symbol; and (iii) domain, which relates firms and individuals to a homepage.  

Because our investment exclusion targets are firms that are either publicly listed or closely related to publicly listed firms, we restricted our prediction targets to firms in the Dow Jones Adverse Media Entity dataset for which we had at least one item of relational information among our prediction targets.  We call the network of our prediction targets the core network.  The core network is a weighted undirected network $G=(V, E, W)$ that consists of a set of nodes V, set of edges E, and edge weights W.  We assume that there is an edge between two nodes in the core network if there is at least one relation type that connects the two nodes.  There are 35,657 firms with 322,138 edges in the core network.  We restrict our attention to the core network because we only have limited information about firms outside this network.  Restricting our focus to the core network strikes a reasonable balance between improving the ``reach''~\cite{Wan2015} of our prediction while assuring that we have sufficient information for prediction.   We also note that the code of the present paper will be made available on the author's website.


\section{Model}
\subsection{Label Propagation Model}

Using the core network defined in the last section, we define a non-negative weight function $f_{\theta}:X \rightarrow [0,1]$, where $X$ defines the set of features for edge $i,j$ extracted from the heterogeneous information network.  We define $f_{\theta}$ to be a simple multilayer perceptron with 30 hidden units and a sigmoid layer for our output function, where $\theta$ denotes the parameters of the model.

We combined the core network defined above with the indicator label of each investment exclusion list category using a variation of label propagation model with edge weight learning using Jacobi iteration~\cite{Chapelle2010}.  Our model is similar in spirit to a supervised random walk~\cite{Backstrom2011}; however, instead of a directed network, we focus on the undirected case.  Our strategy is to split the nodes into the source and target nodes depending on the date of the last negative news report date.  We trained our model to minimize the loss of predicting the labels of our target nodes.  The exact steps connecting $X$, the set of features for edge $i,j$, to the loss is described in algorithm~\ref{alg1}.  Note that our model is not exactly a label propagation model because we set $D_{ii}=\Sigma_{j}1_{ij \in E}$ instead of $D_{ii}=\Sigma_{j}w_{ij}$.  The diagonal dominance condition \cite{Chapelle2010} that ensures that the Jacobi iteration converges still holds because $\Sigma_{j}1_{ij \in E} \geq \Sigma_{j}w_{ij}$, which results from the fact that we defined $0 \leq w_{ij} \leq 1$.  Note that our model is exactly equivalent to the classic label propagation when all $w_{ij}$ equal $1$; however, after learning the edge weights, the spectral radius of $A^{-1}W$ becomes smaller than the usual label propagation, which leads the model to focus on propagating the labels to nearby nodes.

After learning the parameters of the model, we consider both the source and target nodes as known labels and predict the future occurrence of negative news reports, for firms that did not have such news report before in the dataset, after the last date of the training data (i.e., February 1, 2017) to the end of the dataset (i.e., May 31, 2018).  The duration that separates target nodes from source nodes in the training data was set to 31 days before the last date of the training data for most of the investment exclusion list categories for which we had sufficient negative news report information, and 182 days for categories with less information (e.g., sanction, human, and association).  Note that we use the timestamp information to separate the source nodes and target nodes used for training.  More aggressive use of the timestamp information is possible, but this is left for future work. We have also performed a robustness check of our results varying from the last date of the training data to August 1, 2017, and also report results obtained by eliminating the first year (i.e., January 1, 2012, to December 31, 2012) out of the dataset. We obtain very similar results, as shown in the current paper.

We have also performed robustness check of our results varying the last date of the training data to August 1, 2017, and also report results obtained by eliminating the first year (i.e., January 1, 2012, to December 31, 2012) out of the dataset. We obtain similar results as shown in the current paper (this is reported in the supplementary material).



%

\begin{algorithm} 
	\caption{Label propagation with edge weight learning} 
	\label{alg1} 
	\begin{algorithmic}
		\STATE (1) For each edge in the core network set, $w_{ij}=f_{\theta}(x_{ij})$, where $x_{ij}$ denotes features from the network.
		\STATE (2) Compute diagonal degree matrix $D$ using $D_{ii}=\Sigma_{j}1_{ij \in E}$.
		\STATE (3) Compute $A_{ii}=I_{l}(i)+D_{ii}$, where $I_{l}(i)$ indicates $i$'s known label. 
		\STATE (4) Initialize $Y^{0} = (y_{1},...,y_{l},0,...,0)$, where $l$ is the number of known labels.
		\STATE (5) Iterate $Y^{t+1}=A^{-1}(WY^{t}+Y^{0})$ until convergence.
		\STATE (6) Calculate the loss by considering the mean squared error of 	$Y^{target}=(y_{l+1},...,y_{l+m},0,...,0)$ and $Y^{T}=(y_{l+1}^{T},...)$.  
		\STATE (7) Update $\theta$ in $f_{\theta}$ using gradient descent. 
		\STATE (8) Repeat until convergence.
	\end{algorithmic}
\end{algorithm}

\subsection{Edge Features}

For our model to work, we need to define the features for each edge.  We use  the occurrence of relation types in the core network, a path in the overall heterogeneous information network that connect the two nodes~\cite{Lao2011}, or the relation types along path segments that connect the two nodes as our features.  We denote each model as LP-core-relation, LP-path, and LP-path-segment, respectively, where LP denotes ``label propagation.''  Instead of using the raw number counts of each relation type or path, we use a binary indicator to describe whether a specific feature exists.

To be more specific, suppose that edge $A,B$ has the following two direct relations and two paths between them: (A,supplies,B), (A,strategic alliance,B), (A,is in,c,is in,B), and (A,makes,x,is made of,y,makes,B).  For LP-core-relation, we only pay attention to (A, supplies, B) and (A,strategic alliance,B), and hence use $[0,...,1,0,1,0...]$ as our feature, where the two $1$'s correspond to the supplies and strategic alliance relation types.  LP-path works similarly, but instead of creating a one-hot vector for each relation type, we create a one-hot vector for each path.  We restrict our attention to the top 3,000 paths found with a length no larger than 4 for computational reasons.  We also ignore the direction of each relation type.

Moreover, we discard paths that connect two nodes that are already connected by shorter paths. Using our example above, paths with lengths 1 and 2 are not affected by this restriction but, starting from paths of length 3, there might be a path of length 3, such as (A,is in,c,alliance with,d,supports,B), that also connects A and B.  We ignore these paths because node c already appears in a path of length 2 (i.e., (A,is in,c,is in,B)).  We use this additional restriction to prevent super-nodes (e.g., industries) from contaminating our path features.

Features in LP-path-segment are created by distinguishing relation types that occur along the path segments.  This can be considered as a collapsed version of LP-path with relation-type one-hot vectors for each path segment.  A naive implementation of this results in 10 segments for path lengths of up to 4.  However, because the core network is undirected, we can exploit the symmetry and reduce the number of segments.  For example, there is no difference between starting a path from A or starting from B in (A, is in,c, is in, B).  Hence, we do not need to distinguish path segments for paths of length 2, for example, 2:1 and 2:2, but instead we could combine them, thereby creating only one feature of path length 2.  We use path lengths of up to 4, and there are six possible path segments in total, which we denote by 1, 2, 3:1, 3:2, 4:1, and 4:2.

\subsection{Other Models Compared}

We compare our models with the following basic as well as state-of-the-art methods, both using and not fully using the heterogeneous information network.  For the basic model that does not fully use the heterogeneous information network, we add country, industry categories and node degree to Table~\ref{table:adme}, transform the former two into one-hot vectors and use a random forest model for classification.  We call this model the ``random forest.''  For a model that uses the network but not edge weight learning, we directly perform label propagation on the core network.  We call this the ``LP-fixed model.'' We further compare our method with methods that can incorporate multi-category correlation.  Many previous studies have combined multi-category correlation with label propagation~\cite{Wang2016}.  However, most of these methods are computationally very expensive, and hence we use the method of ~\cite{Wang2016}, which turned out to be computationally reasonable.  However, ~\cite{Wang2016} used a KNN graph that is not available in our case.  Instead, we use the core matrix and multiply it by an additional parameter to ensure that the spectral radius of the entire matrix is below one.

\section{Results}

\subsection{Quantitative Comparison}

\begin{figure}[thp]
	\centering
	\includegraphics[width=0.6\linewidth]{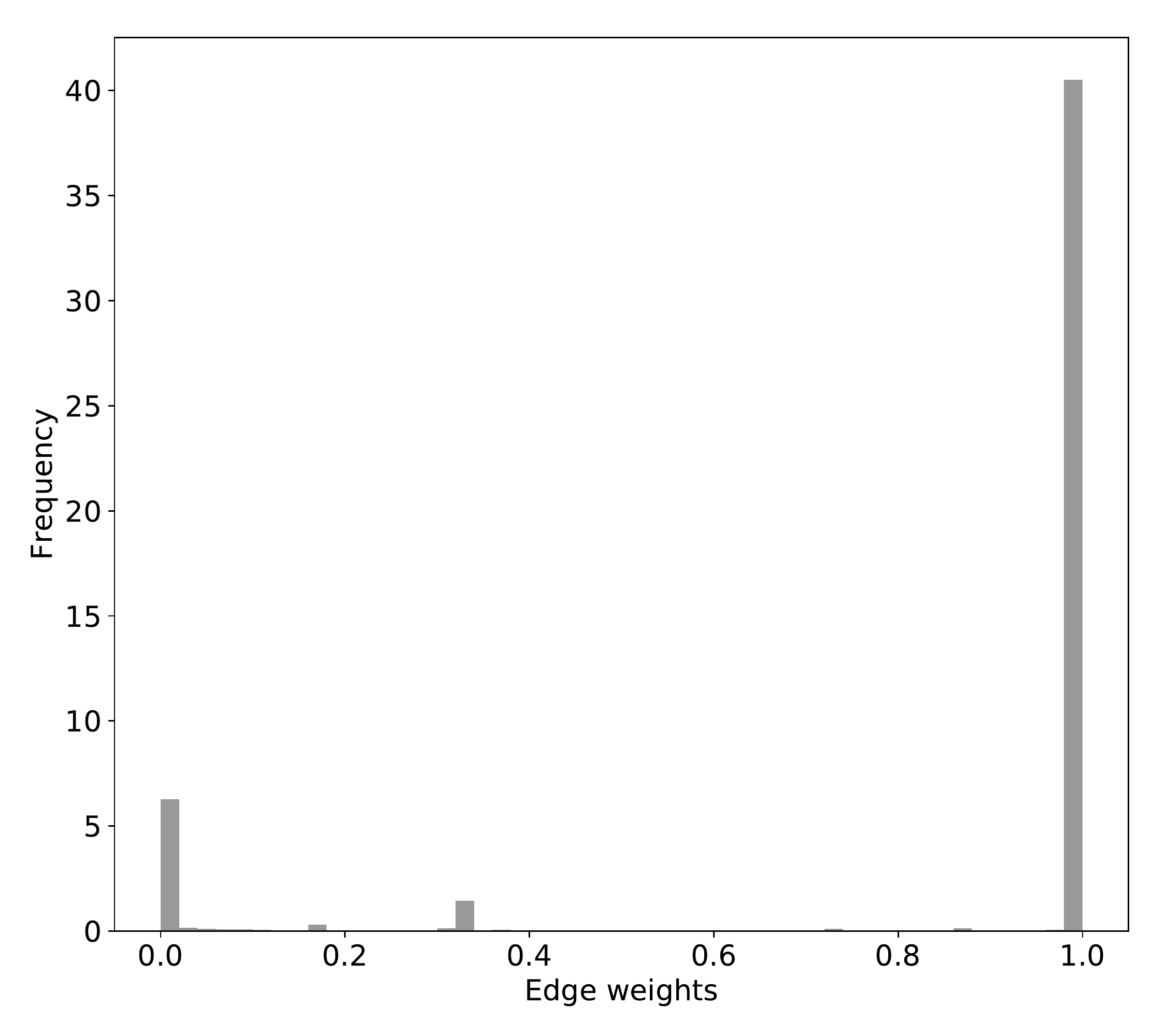}
	\caption{Normalized histogram for the edge weights of the ``Product/Service'' category for LP-path-segment.}
	\label{si:fig:edge_weight}
\end{figure}

\begin{figure*}
	\begin{subfigure}{0.5\textwidth}
		\centering
		\includegraphics[width=1.0\linewidth]{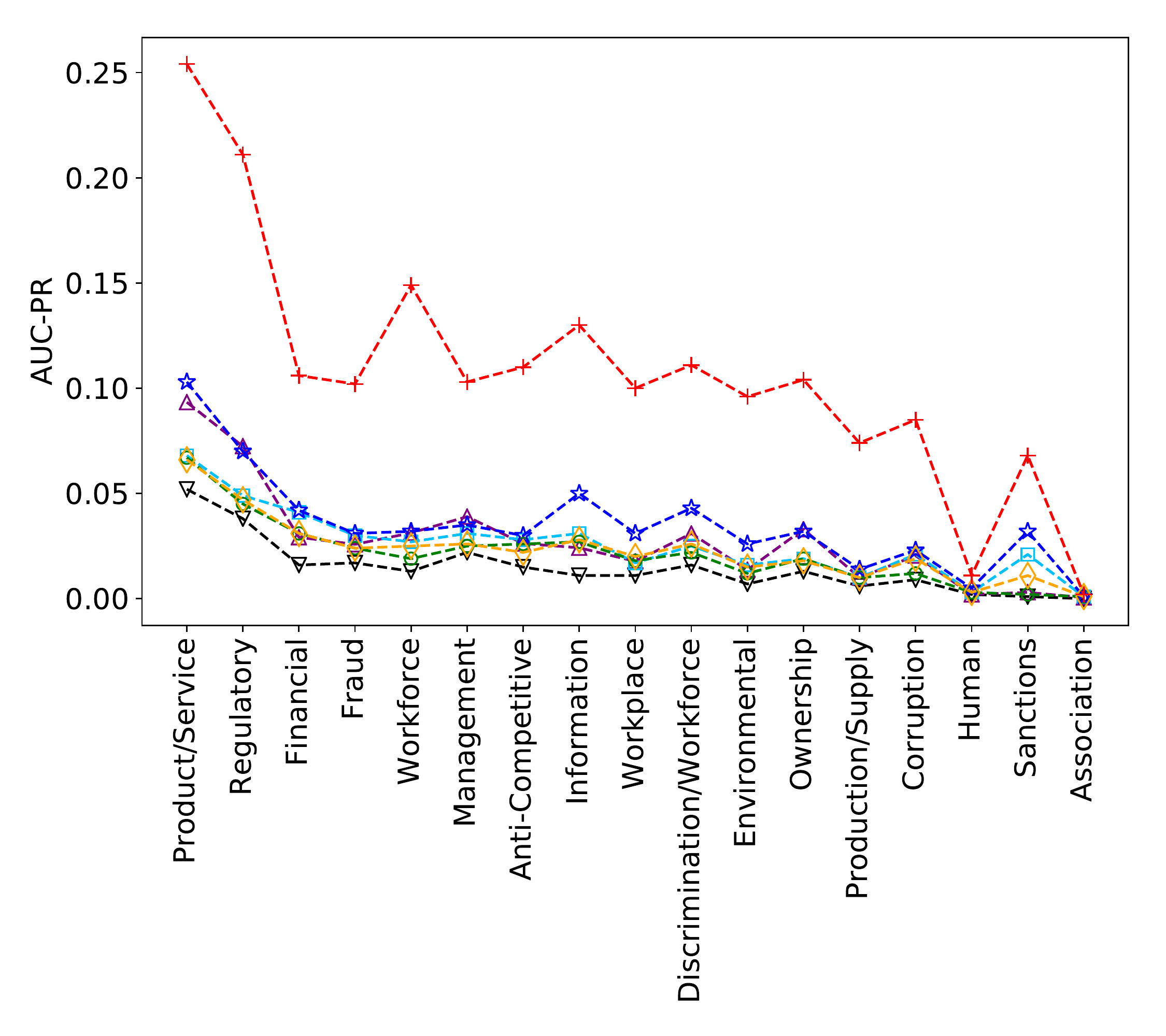}
		\caption{AUC-PR}
		\label{fig:aucpr}
	\end{subfigure}%
	\begin{subfigure}{0.5\textwidth}
		\centering
		\includegraphics[width=1.0\linewidth]{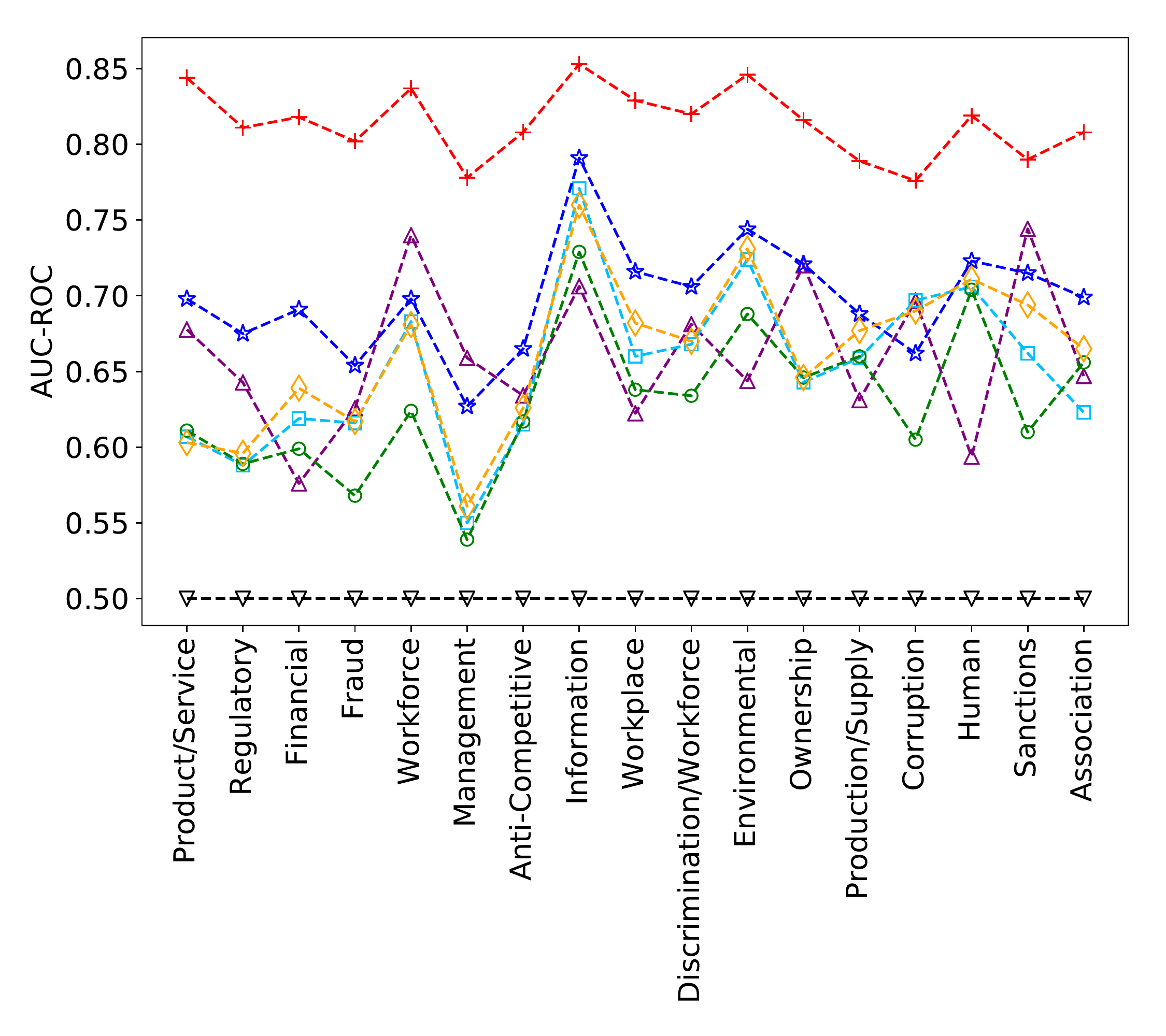}
		\caption{AUC-ROC}
		\label{fig:aucroc}
	\end{subfigure}
	\caption{Comparison of predictive performance for random guessing (black inverted triangles), random forest (purple triangles), LP-fixed (light-blue squares), LP-mult (green circles), LP-core-relation (blue stars), LP-path (orange diamonds), and LP-path-segment (red crosses).}
	\label{fig:perform}
\end{figure*}

Our prediction problem is a standard binary classification problem (whether a firm would be added to the investment exclusion list from February 1, 2017, to May 31, 2018), so we use the area under the receiver operator characteristics (AUC-ROC) for evaluation.  Because our labels are highly imbalanced, we also evaluate performance using the area under the precision-recall curves (AUC-PR)~\cite{Davis2006}.  Because of space limitations, the results are shown in the form of graphics (see Fig.~\ref{fig:perform}).

We first note that there seems to be predictability by only performing label propagation on the core network (i.e., LP-fixed).  However, its performance is slightly worse than that of the random forest baseline using country and industry indicators.  The performance of the network approach improves when the adaptive edge weighting scheme is used.  This is apparent because LP-core-relation performs better than LP-fixed almost all the time.  It is possible that LP-path performs worse than LP-core-relation because we only use the top 3,000 paths for computational reasons.  LP-mult does not seem to improve performance when compared with LP-fixed.  Whether this originates from the particular algorithm used or because not much information is added by incorporating multi-category correlation needs further investigation.  Finally, comparing LP-path-segment to all the other methods, we find that it performs substantially better, outperforming all the methods for all the categories compared in this paper.  To summarize, our results show that using the information stored in the heterogeneous information network leads to a substantially better predictive accuracy.

For completeness, in Fig.~\ref{si:fig:edge_weight}, we provide a normalized histogram that shows the learned edge weights for LP-path-segment for predicting the ``Product/Service'' category.  We see that our algorithm tends to separate edge weights into values of either one or zero.

\subsection{Interpretability}

To understand what our models have learned, we perform partial dependence analysis on our learned model~\cite{Hastie2001}.  However, because the features used by LP-path-segment are highly correlated, calculating the importance measure for each feature might not be a reasonable approach.  Hence, we first reduce the dimensionality of the feature space to 50 using a standard binary nonnegative matrix factorization (BNMF) technique \cite{Zitnik2012} and then perform the usual partial dependence analysis along the basis of the matrix obtained by the standard BNMF method.  The BNMF finds similar relation types among the different path segments that can be aligned to make an interpretation of the results possible.  Typically, the sample standard deviation of the fitted values of the partial dependence plot is used as a  measure of feature importance~\cite{Greenwell2017}.  However, because our feature matrix is binary, we instead focus on the absolute difference of the response at the 0.99 and 0.01 quantile of the coefficient vector that corresponds to each basis vector.  We also consider the average value of the importance measure, repeating the training and partial dependence analysis step 30 times using different initial parameters to mitigate the effect of fluctuation that results from the learning process.

\begin{figure*}
	\centering
	\begin{subfigure}{0.5\textwidth}
		\centering
		\includegraphics[width=.95\linewidth]{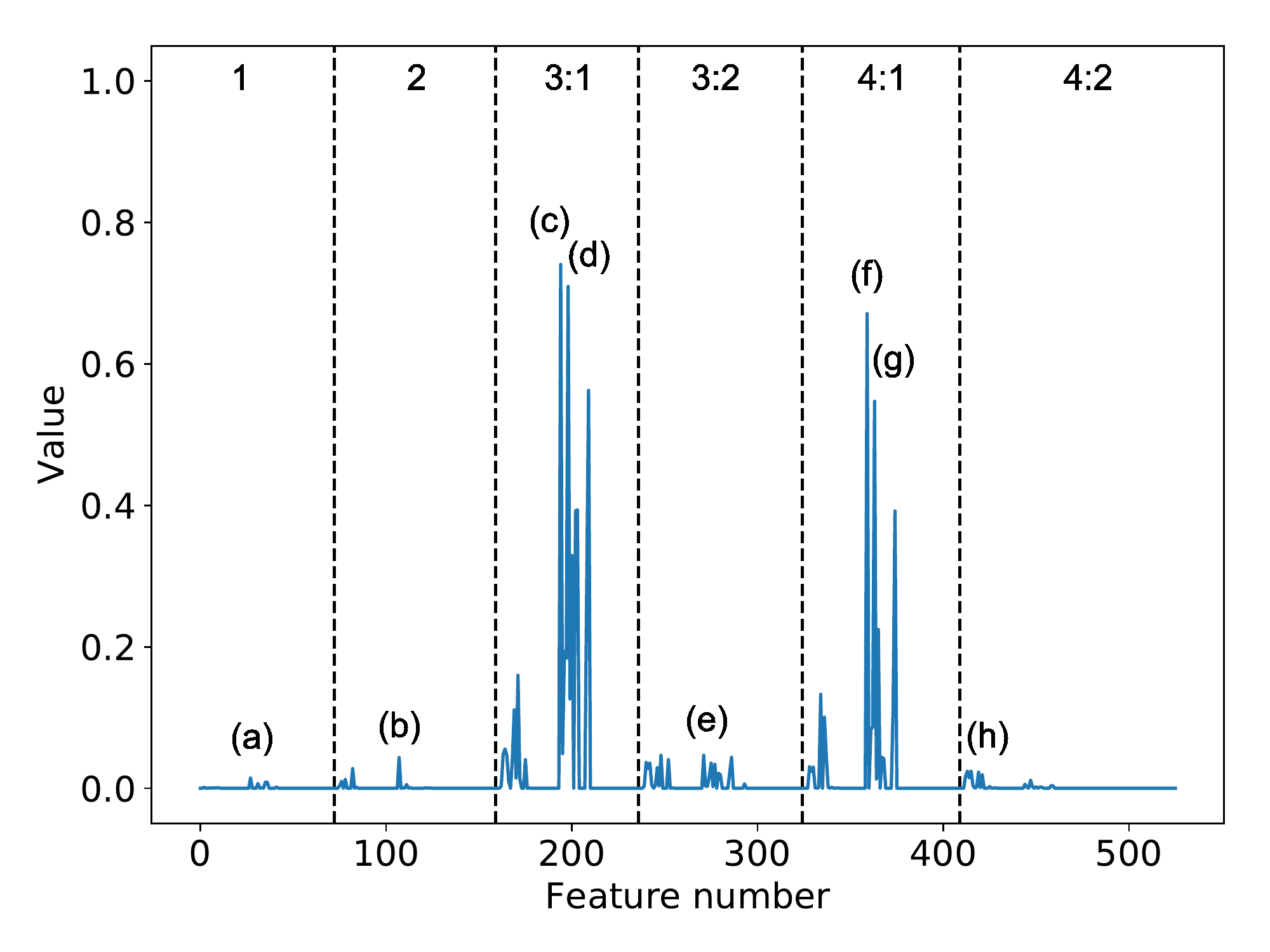}
		\caption{Basis vector 4 (license-licensee)}
		\label{fig:sfig1}
	\end{subfigure}%
	\begin{subfigure}{0.5\textwidth}
		\centering
		\includegraphics[width=.95\linewidth]{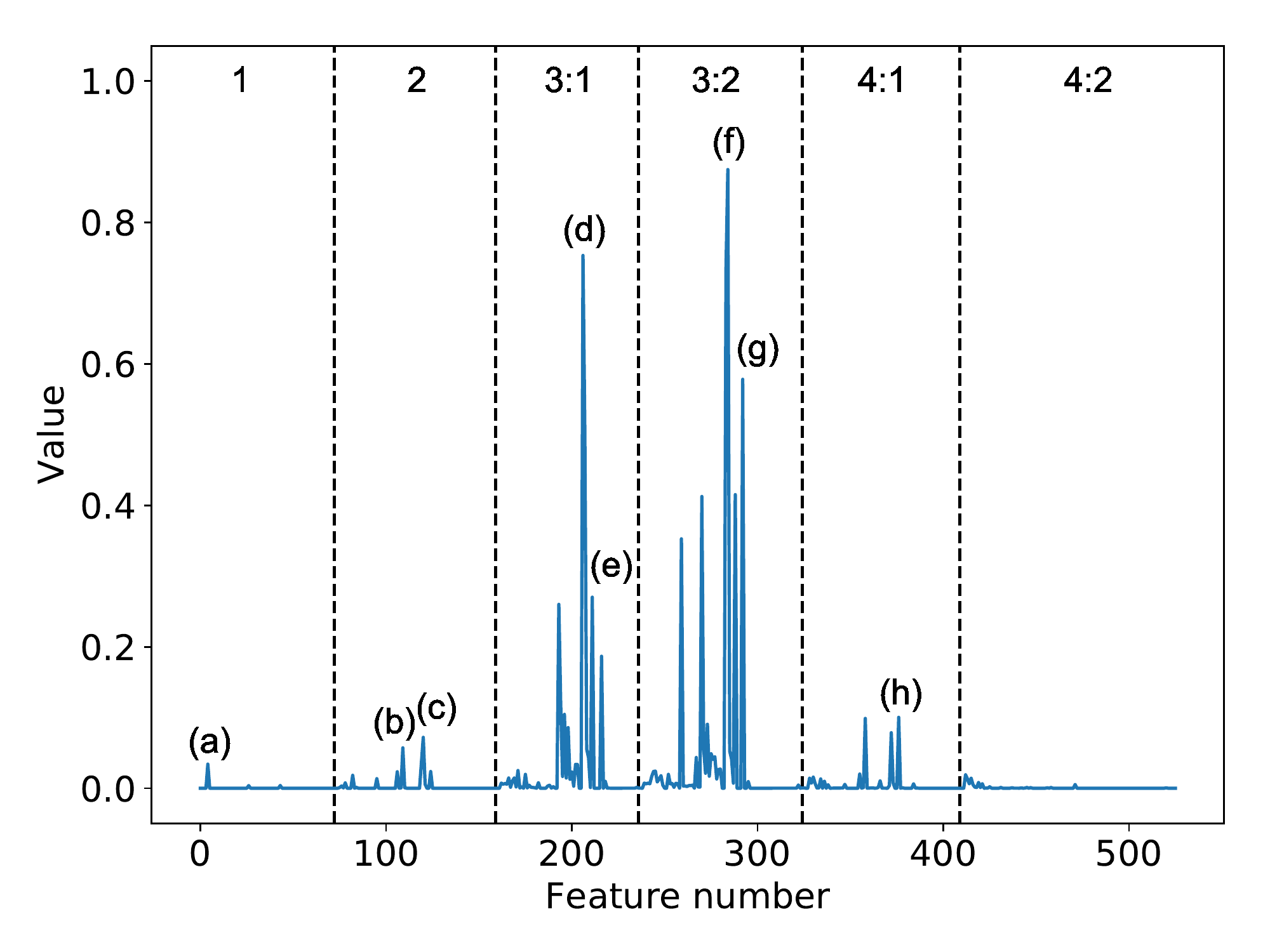}
		\caption{Basis vector 13 (buyer-seller)}
		\label{fig:sfig2}
	\end{subfigure}
	\caption{Comparison of basis vector 4 and basis vector 13.  The dotted vertical lines divide each path segment.  Because there are relation types that do not appear in some path segments, the total number of features is 526 instead of 1,296 ($216 \times 6$).  Peaks in basis vector 4: (a) in-licensing, (b) in-licensing,  (c) in-licensing, (d) out-licensing, (e) distributor, (f) in-licensing, (g) out-licensing, and (h) customer.  Peaks in basis vector 13: (a) customer, (b) partner-manufacture, (c) international shipping (d) receive goods, (e) international shipping, (f) international shipping (g) receive goods,  and (h) franchise.}
	\label{fig:pdp:product}
\end{figure*}

Table~\ref{table:topfeature} shows the top five important features learned for the ``Product/Service'' category.  Basis vector 4 seems to have the most negative effect, whereas basis vector 13 seems to have the most positive effect on the weights.  Note that features in higher path segments are likely to have a higher value in the basis vector because our feature matrix is a binary matrix taking one if there is at least one relation type in a particular path segment.  Thus, we must pay attention to the relation type in each segment when interpreting the result and, in Fig.~\ref{fig:pdp:product}, we report the top relation types for each path segment for basis vector 4 and basis vector 13.   Whereas the path segments of basis vector 4 include more relation types that are related to the license relation, basis vector 13, which has a positive effect, focuses more on the buyer-seller and partnership-manufacture relations.  Because ``Product/Service'' is more closely related to news about the specific products of a firm, such as recall incidents and drug test failures,  our model learned to value those relation types in the path segments more.

\begin{figure*}
	\begin{subfigure}{.5\textwidth}
		\centering
		\includegraphics[width=.95\linewidth]{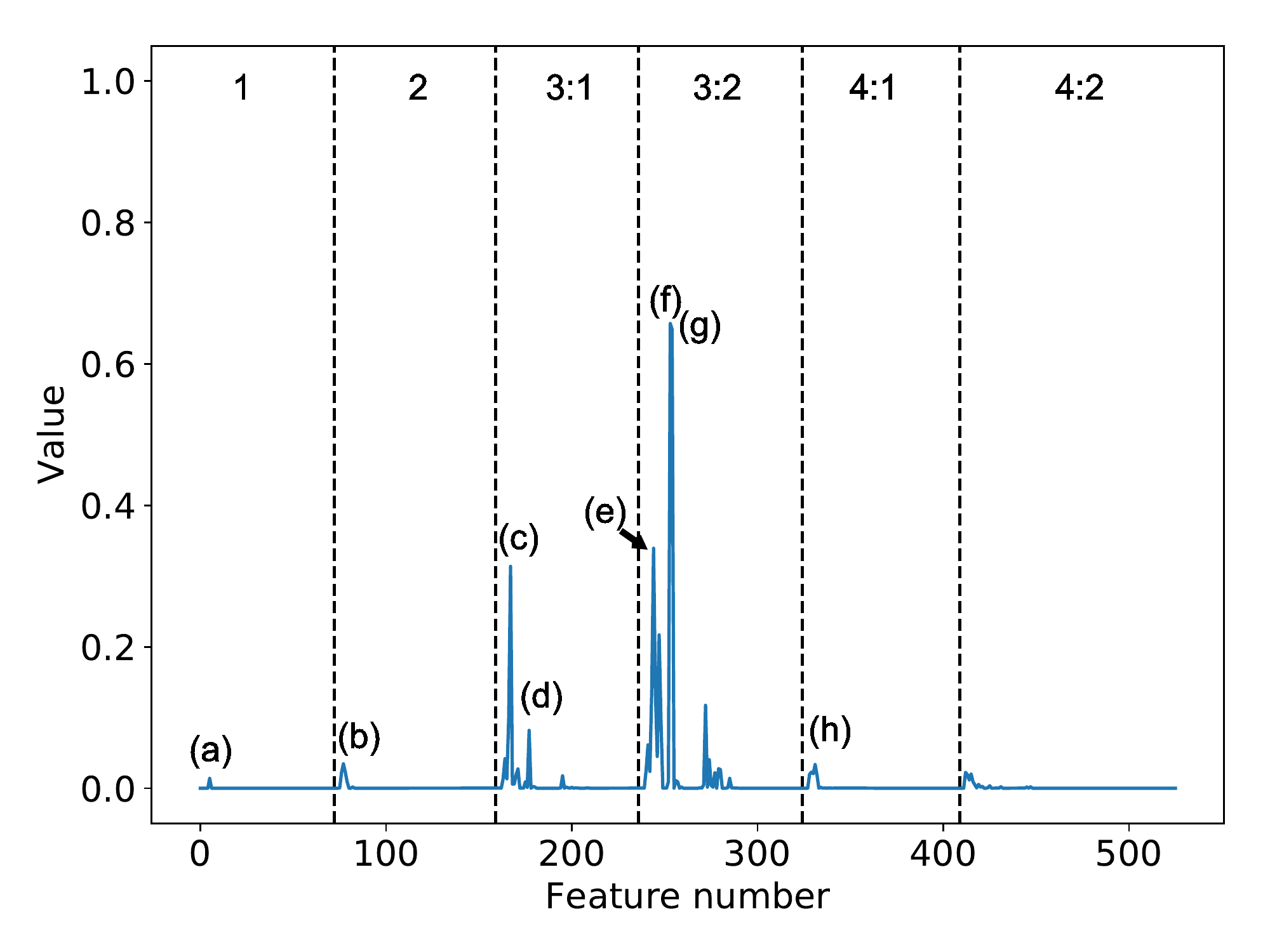}
		\caption{Basis vector 34 (creditor borrower)}
		\label{fig:pdp34}
	\end{subfigure}%
	\begin{subfigure}{.5\textwidth}
		\centering
		\includegraphics[width=.95\linewidth]{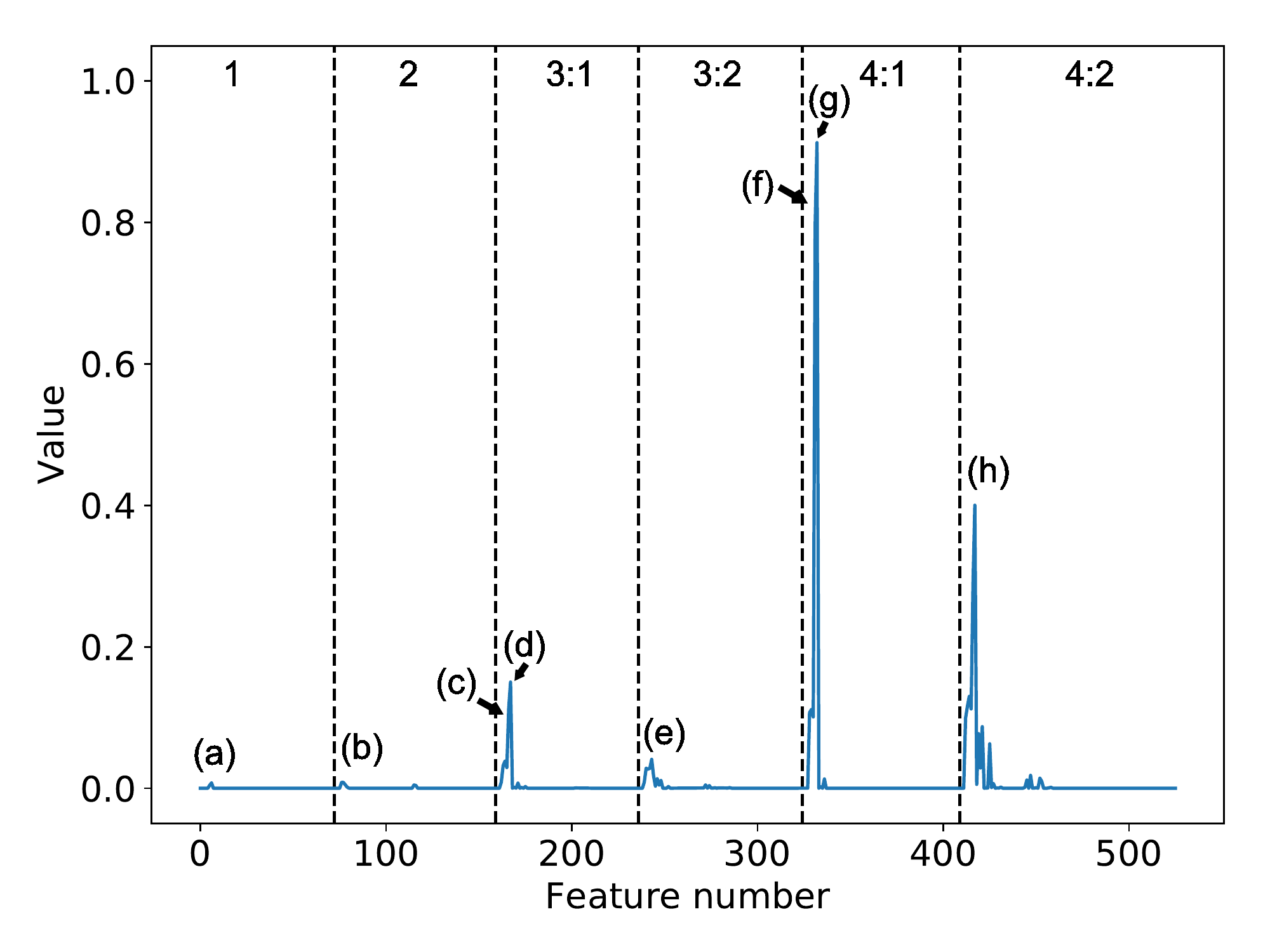}
		\caption{Basis vector 10 (creditor borrower)}
		\label{fig:pdp10}
	\end{subfigure}
	\caption{Comparison of basis vector 34 and basis vector 10.  The dotted vertical lines divide each path segment.  Because there are relation types that do not appear in some path segments, the total number of features is 526 instead of 1,296 ($216 \times 6$).  Peaks in basis vector 34: (a) creditor, (b) strategic alliance, (c) borrower, (d) creditor, (e) borrower, (f) tenant, (g) landlord, and (h) creditor.  Peaks in basis vector 10: (a) borrower, (b) strategic alliance, (c) creditor, (d) borrower, (e) creditor, (f) creditor, (g) borrower, and (h) borrower.}
	\label{fig:pdp:financial}
\end{figure*}

In Table~\ref{si:table:topfeature:financial}, we show the top five important features for the ``Financial'' category.  All the top five features have a positive effect on the edge weights, so we focus on the top two and report analysis for basis vector 34 (Fig.~\ref{fig:pdp34}) and basis vector 10 (Fig.~\ref{fig:pdp10}).  For basis vectors 34 and 10, we see that they focus more on creditor-borrower relationships.  Because ``Financial'' negative news is reported when a firm is in a serious financial condition or when there are ownership issues, it makes sense that these relation types are at the top and have a positive effect on the edge weights.

Since reporting the names of our prediction might be too offensive, we refrain from doing that in the present paper, but we have also checked several examples from our prediction and checked the validity of our approach as well.

\begin{table}
	\centering
	\begin{tabular}{lrrr}
		\toprule
		Rank & Basis & $E_{\hat{\theta} }[f(x_{0.99})-f(x_{0.01})]$ & $|E_{\hat{\theta} }[f(x_{0.99})-f(x_{0.01})]|$\\
		\midrule
		1 & 4 &  -0.096 & 0.096\\
		2 & 26 & -0.070 & 0.070\\
		3 & 30 & -0.057 & 0.057\\
		4 & 13 & 0.040 & 0.040\\
		5 & 7 &  0.039 & 0.039\\
		\bottomrule
	\end{tabular}
	\caption{Top five important features for the ``Product/Service'' category.}
	\label{table:topfeature}
\end{table}

\begin{table}[tbhp]
	\centering
	\begin{tabular}{lrrr}
		\toprule
		Rank & Basis & $E_{\hat{\theta} }[f(x_{0.99})-f(x_{0.01})]$ & $|E_{\hat{\theta} }[f(x_{0.99})-f(x_{0.01})]|$\\
		\midrule
		1 & 34 &  0.090 & 0.090 \\
		2 & 10 & 0.089 & 0.089\\
		3 & 7 & 0.089 & 0.089 \\		
		4 & 21 & 0.088 & 0.088\\
		5 & 20 &  0.081 & 0.081\\
		\bottomrule
	\end{tabular}
	\caption{Top five important features for the ``Financial'' category.  }
	\label{si:table:topfeature:financial}
\end{table}

\section{Conclusion}

In this paper, using a comprehensive dataset of negative news investment exclusion list data and a heterogeneous information network among 35,657 global firms assembled from professional data sources, we showed that the predictive performance of predicting firms that are more likely to be added to an investment exclusion list increases in a striking manner when we exploit the vast amount of information stored in the heterogeneous information network.  Our work suggests a machine-assisted method to exploit the heterogeneous information contained in big data to monitor firms on a global scale for better risk management.  We also showed that our model is interpretable.

Fig.~\ref{fig:perform} demonstrates the remarkable over-performance of our methods, which requires some explanation. First, when a problem occurs for a firm, it is likely that the firms that it is related to or similar firms are also in trouble.  The similarity of firms could be quantified by the closeness in the heterogeneous information network, which includes a variety of information concerning a firm.  Moreover, instead of using the raw closeness measure that our heterogeneous information network suggests, we adjust for the closeness measure using past patterns, which results in high predictive performance.  Perhaps more importantly, when a problem catches the eye of the public, investigative journalists search for nearby firms for follow-up stories.  By doing so, they can claim that the first problem they reported is not just confined to one firm, but a more general issue in need of more attention.   Hence, it might not be surprising that our machine-assisted method works.

The misclassifications of our model can be organized into four categories, as shown in Table~\ref{table:error}.  The inaccuracy that results from our model or data limitations could result in both false positive and false negative errors.  There are exogenous events in false negatives that are impossible to predict from our approach of simply learning past negative news patterns. Exogenous events always constitute an intrinsic limit to prediction methods. However, on the positive side, there might be cases of false positive misclassifications that correspond to unrealized or uncovered events.  From a journalist's point of view, the list of firms in this category might be the next possible target for further investigation.  From a firm's point of view, our prediction score might be a good diagnostic to follow to take timely actions for fair media coverage using firm-initiated press releases and investor relations firms \cite{Solomon2012}. Moreover, instead of using the media labels as the data vendor provides it, we could investigate further into the text to pick up news that had a significant impact (e.g., arrest, lawsuits) instead of just a shallow allegation.  We could also take into account node information (e.g., firm size) to focus on firms that are too big to fail or the banking sector for which the effect of negatvie media coverage is already well-known \cite{Birchler2016}.  

\begin{table}[H]
	\resizebox{1.0\textwidth}{!}{
		\centering
		\begin{tabular}{lrrr}
			\toprule
			{}     &  {}   & \multicolumn{2}{c}{Real world}\\
			\multicolumn{2}{c}{} & False & True\\
			\midrule
			Prediction & False &  Correct                  & FN: Model error/Data limit\\
			{}    & {}    & {}                        & Exogeneous events\\
			{}    &  True &  FP: Model error/Data limit   & Correct \\
			{}    &  {}   &  Not realized/Not covered & {} \\
			\bottomrule
	\end{tabular}}
	\caption{Model prediction and the real world. FP denotes false positive and FN denotes false negative.}
	\label{table:error}
\end{table}

\bibliographystyle{plain}
\bibliography{Hisano_Bib}

\begin{thebibliography}{10}

\bibitem{Auer2007}
S\"{o}ren Auer, Christian Bizer, Georgi Kobilarov, Jens Lehmann, Richard
  Cyganiak, and Zachary Ives.
\newblock Dbpedia: A nucleus for a web of open data.
\newblock In {\em Proceedings of the 6th International The Semantic Web and 2Nd
  Asian Conference on Asian Semantic Web Conference}, ISWC'07/ASWC'07, pages
  722--735, Berlin, Heidelberg, 2007. Springer-Verlag.

\bibitem{Backstrom2011}
Lars Backstrom and Jure Leskovec.
\newblock Supervised random walks: Predicting and recommending links in social
  networks.
\newblock In {\em Proceedings of the Fourth ACM International Conference on Web
  Search and Data Mining}, WSDM '11, pages 635--644, New York, NY, USA, 2011.
  ACM.

\bibitem{Birchler2016}
Urs Birchler, René Hegglin, Michael~R. Reichenecker, and Alexander~F. Wagner.
\newblock Which swiss gnomes attract money? efficiency and reputation as
  performance drivers of wealth management banks.
\newblock {\em Swiss Finance Institute Research Paper}, No. 16-28, 2016.

\bibitem{Cao2018}
B.~Cao, M.~Mao, S.~Viidu, and P.~S. Yu.
\newblock Hitfraud: A broad learning approach for collective fraud detection in
  heterogeneous information networks.
\newblock In {\em 2017 IEEE International Conference on Data Mining (ICDM)},
  volume~00, pages 769--774, Nov. 2018.

\bibitem{Chapelle2010}
Olivier Chapelle, Bernhard Schlkopf, and Alexander Zien.
\newblock {\em Semi-Supervised Learning}.
\newblock The MIT Press, 1st edition, 2010.

\bibitem{Chen2016}
Y.~Chen, R.~Liu, and W.~Xu.
\newblock Movie recommendation in heterogeneous information networks.
\newblock In {\em 2016 IEEE Information Technology, Networking, Electronic and
  Automation Control Conference}, pages 637--640, May 2016.

\bibitem{Davis2006}
Jesse Davis and Mark Goadrich.
\newblock The relationship between precision-recall and roc curves.
\newblock In {\em Proceedings of the 23rd International Conference on Machine
  Learning}, ICML '06, pages 233--240, New York, NY, USA, 2006. ACM.

\bibitem{Dong2014}
Xin Dong, Evgeniy Gabrilovich, Geremy Heitz, Wilko Horn, Ni~Lao, Kevin Murphy,
  Thomas Strohmann, Shaohua Sun, and Wei Zhang.
\newblock Knowledge vault: A web-scale approach to probabilistic knowledge
  fusion.
\newblock In {\em Proceedings of the 20th ACM SIGKDD International Conference
  on Knowledge Discovery and Data Mining}, KDD '14, pages 601--610, New York,
  NY, USA, 2014. ACM.

\bibitem{Furukawa2017}
Katsuhisa Furukawa.
\newblock {\em Kitacyosen Kaku no Shikingen Kokuren Sousa no Hiroku [Funding
  Source of North Korea: A Note on United Nation's Investigation] Funding
  source}.
\newblock Tokyo Shincyosya, Tokyo, Japan, 2017.

\bibitem{Greenwell2017}
Brandon~M. Greenwell.
\newblock pdp: An r package for constructing partial dependence plots.
\newblock {\em The R Journal}, 9(1):421--436, 2017.

\bibitem{Hastie2001}
Trevor Hastie, Robert Tibshirani, and Jerome Friedman.
\newblock {\em The Elements of Statistical Learning}.
\newblock Springer Series in Statistics. Springer New York Inc., New York, NY,
  USA, 2001.

\bibitem{Hill2010}
C.~Hill.
\newblock A survey of heterogeneous information network analysis.
\newblock {\em U. Pitt. L. Rev.}, 585, 2010.

\bibitem{Hofmann2017}
Alexandra Hofmann, Samresh Perchani, Jan Portisch, Sven Hertling, and Heiko
  Paulheim.
\newblock Dbkwik: towards knowledge graph creation from thousands of wikis.
\newblock In {\em ISWC-P\&D-Industry 2017 : Proceedings of the ISWC 2017
  Posters \& Demonstrations and Industry Tracks co-located with 16th
  International Semantic Web Conference (ISWC 2017) Vienna, Austria, October
  23rd to 25th, 2017}, volume 1963, page Paper 540, Aachen, 2017. RWTH.

\bibitem{Hu2018}
Yujing Hu, Qing Da, Anxiang Zeng, Yang Yu, and Yinghui Xu.
\newblock Reinforcement learning to rank in e-commerce search engine:
  Formalization, analysis, and application.
\newblock In {\em Proceedings of the 24th ACM SIGKDD International Conference
  on Knowledge Discovery \&\#38; Data Mining}, KDD '18, pages 368--377, New
  York, NY, USA, 2018. ACM.

\bibitem{Lao2011}
Ni~Lao, Tom Mitchell, and William~W. Cohen.
\newblock Random walk inference and learning in a large scale knowledge base.
\newblock In {\em Proceedings of the Conference on Empirical Methods in Natural
  Language Processing}, EMNLP '11, pages 529--539, Stroudsburg, PA, USA, 2011.
  Association for Computational Linguistics.

\bibitem{Markham2006}
J.W. Markham.
\newblock {\em A Financial History of Modern U.S. Corporate Scandals: From
  Enron to Reform}.
\newblock M.E. Sharpe, 2006.

\bibitem{Nickel2016}
Maximilian Nickel, Kevin Murphy, Volker Tresp, and Evgeniy Gabrilovich.
\newblock A review of relational machine learning for knowledge graphs.
\newblock {\em Proceedings of the {IEEE}}, 104(1):11--33, 2016.

\bibitem{OECD2017}
OECD.
\newblock Responsible business conduct for institutional investors: Key
  considerations for due diligence under the oecd guidelines for multinational
  enterprises.
\newblock {\em OECD guidlines}, 2017.

\bibitem{WFE2017}
World~Federation of~Exchanges.
\newblock Wfe annual statistics guide 2017.
\newblock 2017.

\bibitem{Solomon2012}
David~H. Solomon.
\newblock {Selective Publicity and Stock Prices}.
\newblock {\em Journal of Finance}, 67(2):599--638, April 2012.

\bibitem{Sun2013}
Yizhou Sun and Jiawei Han.
\newblock Mining heterogeneous information networks: A structural analysis
  approach.
\newblock {\em SIGKDD Explor. Newsl.}, 14(2):20--28, April 2013.

\bibitem{Zitnik2012}
Marinka \v{Z}itnik and Bla\v{z} Zupan.
\newblock Nimfa: A python library for nonnegative matrix factorization.
\newblock {\em J. Mach. Learn. Res.}, 13(1):849--853, March 2012.

\bibitem{Sherwood2018}
Matthew W.~Sherwood and Julia Pollard.
\newblock {\em Responsible Investing: An Introduction to Environmental, Social,
  and Governance Investments}.
\newblock Routledge, 09 2018.

\bibitem{Wan2015}
Chang Wan, Xiang Li, Ben Kao, Xiao Yu, Quanquan Gu, David Cheung, and Jiawei
  Han.
\newblock Classification with active learning and meta-paths in heterogeneous
  information networks.
\newblock In {\em Proceedings of the 24th ACM International on Conference on
  Information and Knowledge Management}, CIKM '15, pages 443--452, New York,
  NY, USA, 2015. ACM.

\bibitem{Wang2016}
Daixin Wang, Peng Cui, and Wenwu Zhu.
\newblock Structural deep network embedding.
\newblock In {\em Proceedings of the 22Nd ACM SIGKDD International Conference
  on Knowledge Discovery and Data Mining}, KDD '16, pages 1225--1234, New York,
  NY, USA, 2016. ACM.

\bibitem{Wang2018}
Hongwei Wang, Fuzheng Zhang, Jialin Wang, Miao Zhao, Wenjie Li, Xing Xie, and
  Minyi Guo.
\newblock Ripplenet: Propagating user preferences on the knowledge graph for
  recommender systems.
\newblock In {\em Proceedings of the 27th ACM International Conference on
  Information and Knowledge Management}, CIKM '18, pages 417--426, New York,
  NY, USA, 2018. ACM.

\bibitem{Wang2017Review}
Q.~Wang, Z.~Mao, B.~Wang, and L.~Guo.
\newblock Knowledge graph embedding: A survey of approaches and applications.
\newblock {\em IEEE Transactions on Knowledge and Data Engineering},
  29(12):2724--2743, Dec 2017.

\bibitem{Wilcke2017}
Xander Wilcke, Peter Bloem, and Victor {De Boer}.
\newblock The knowledge graph as the default data model for learning on
  heterogeneous knowledge.
\newblock {\em Data Science}, 1(1-2):39--57, 12 2017.

\end{thebibliography}

\end{document}